\documentclass[twocolumn,aps,prl,floatfix]{revtex4-1}
\usepackage{amsmath}
\usepackage{amsfonts}
\usepackage{amssymb}
\usepackage{graphicx}
\usepackage{bm}
\usepackage{bbold}
\usepackage{float}
\usepackage{hyperref}
\usepackage{color}
\usepackage{tabularx}
\usepackage{url}

\begin{document}
\title{Second-order Topological Superconductors with Mixed Pairing}
\author{Xiaoyu Zhu}
\affiliation{School of Science, MOE Key Laboratory for Non-equilibrium Synthesis and Modulation of Condensed Matter, Xi'an Jiaotong University, Xi'an, Shaanxi 710049, China}

\date{\today}

\begin{abstract}
We show that a two-dimensional semiconductor with Rashba spin-orbit coupling could be driven into the second-order topological superconducting phase when a mixed-pairing state is introduced. The superconducting order we consider involves only even-parity components and meanwhile breaks time-reversal symmetry. As a result, each corner of a square-shaped Rashba semiconductor would host one single Majorana zero mode in the second-order nontrivial phase. Starting from edge physics, we are able to determine the phase boundaries accurately. A simple criterion for the second-order phase is further established, which concerns the relative position between Fermi surfaces and nodal points of the superconducting order parameter. In the end, we propose two setups that may bring this mixed-pairing state into the Rashba semiconductor, followed by a brief discussion on the experimental feasibility of the two platforms.
\end{abstract}

\maketitle

Topological superconductors (TSCs) distinguish themselves from trivial ones in the robust midgap states---Majorana zero modes (MZMs)---that could form either at local defects or boundaries \cite{Read2000,Kitaev2001,Wilczek2009,Qi2011,Alicea2012,Beenakker2013,Stanescu2013,Elliott2015,Aguado2017}. Among the various proposals for TSCs, semiconducting systems with Rashba spin-orbit coupling (RSOC) \cite{Sau2010,Alicea2010,Lutchyn2010,Oreg2010} as well as topological insulating systems \cite{Fu2008} have attracted the most attention. In both platforms, signatures of MZMs have been observed when conventional $s$-wave pairing is introduced through proximity effect \cite{Mourik2012,Das2012,Deng2012,Rokhinson2012,Finck2013,Deng2016,Perge2014,Hart2014,Xu2015,Sun2016,He2017}.

In these conventional, also termed as first-order, TSCs, topologically nontrivial bulk in $d$ dimensions is usually accompanied by MZMs confined at $(d-1)$-dimensional boundaries, the so-called bulk-boundary correspondence. Very recently, this correspondence was extended in topological phases of $n$th order \cite{Benalcazar2017,Benalcazar2017prb,Schindler2018,Song2017,Langbehn2017,Ezawa2018,Khalaf2018,Geier2018,Zhu2018,You2018,Volpez2018,Serra-Garcia2018,Peterson2018,Imhof2018,Noh2018,Schindler2018sa,Slager2015,Dumitru2019,Wang2018,Wu2018}, where topologically protected gapless modes emerge at $(d-n)$-dimensional boundaries. In Refs.[\onlinecite{Yan2018,WangQ2018,Liu2018,ZhangRX2018}], the authors demonstrate that a topological insulator could be transformed into a second-order TSC when unconventional pairing with the $s_\pm$- or $d_{x^2-y^2}$-wave form is introduced. Looking back at the history of first-order TSCs, one may then ask if it is possible for a Rashba semiconductor (RS), which is itself a trivial system as opposed to topological insulators, to accommodate such a higher-order nontrivial phase as well. In this work, we will show that it is possible, provided a mixed-pairing state that exhibits both extended $s$-wave and $d_{x^2-y^2}$-wave symmetries could be induced therein.

Admixture of the two aforementioned pairing states was envisioned shortly after the discovery of iron-based superconductors (FeSCs) \cite{Lee2009,Platt2012,Khodas2012,Fernandes2013}. Since then tremendous efforts have been made to identify this mixed-pairing order \cite{Chubukov2015,Fernandes2017}. In this Letter, we shall consider a general mixed state that could reduce to three intensively studied mixed pairings in FeSCs, that is, $s+d$ \cite{Livanas2015}, $s+is$ \cite{Maiti2013} and $s+id$ \cite{Lee2009,Khodas2012}. Our main finding is that, such a pairing state alone could possibly drive a two-dimensional RS into the second-order topological superconducting phase. Of the three specific forms aforementioned, however, only $s+id$ pairing could make it. An accurate criterion is further established for the second-order phase to emerge, which is closely related to the relative position between nodal points of the pairing order parameter and the two nondegenerate Fermi surfaces split by RSOC.

\begin{figure}[t]
\includegraphics[scale=0.45]{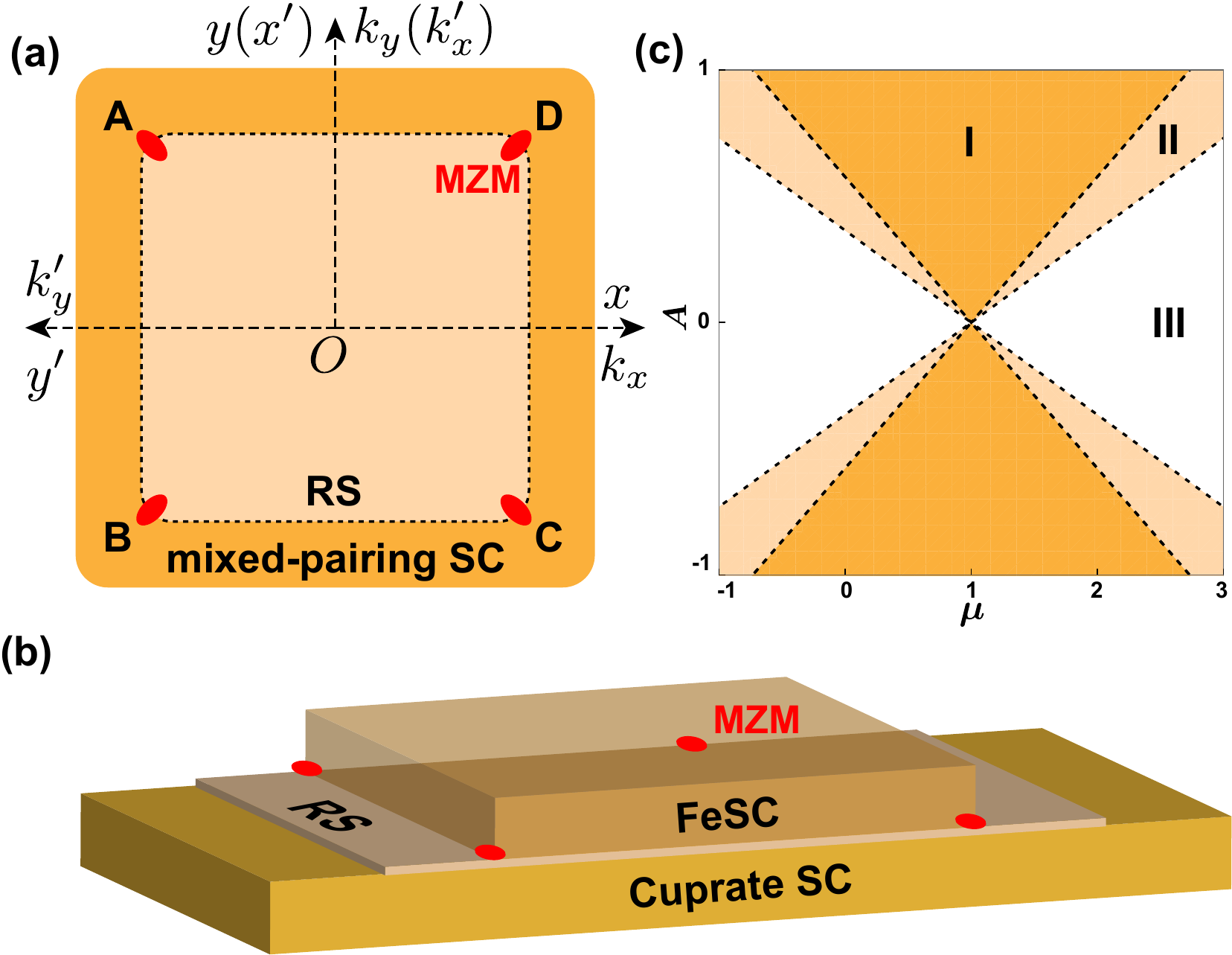}
\caption{(a). Heterostructure of a RS and SC with mixed pairing (top view). Majorana zero modes (denoted by red solid ellipses) emerge at the four corners in the second-order topological superconducting phase. Two sets of coordinate systems connected by $\mathcal C_4$ rotation are shown, in both real space and reciprocal space. (b). Hybrid Josephson junction with FeSC ($s_\pm$ pairing), cuprate SC ($d_{x^2-y^2}$ pairing), and a single RS layer sandwiched between them. The junction interface is parallel to the $ab$-plane of the two SCs. (c). Phase diagram in absence of $\Delta_{sd}$, with $\Delta_0 = \Delta_1 = 1$. Phase I: first-order TSC; Phase II: nodal SC; Phase III: fully gapped trivial SC. For $s+id$ pairing, Phase I and II would be driven into the second-order phase. All parameters in this and the following figures are in the unit of $t$.}\label{fig1}
\end{figure}

We consider a RS in two dimensions with mixed pairing of extended $s$-wave and $d_{x^2-y^2}$-wave form, and the corresponding Hamiltonian is given by
\begin{eqnarray}
&&H =\frac{1}{2} \sum\limits_{\bm k} \Psi^\dagger(\bm k) \mathcal H(\bm k)  \Psi(\bm k), \nonumber\\
&&\mathcal H(\bm k) = h(\bm k)\tau_3+\Delta_s(\bm k)\tau_1+\Delta_{sd}(\bm k)\tau_2,\label{eq:H}
\end{eqnarray}
in the Nambu spinor basis $\Psi(\bm k) = \{c_{\bm k\uparrow}, c_{\bm k\downarrow}, c_{\bm k\downarrow}^\dagger,-c_{\bm k\uparrow}^\dagger\}^T$. In Eq. (\ref{eq:H}), $h(\bm k) = 2A(\sin k_x \sigma_2-\sin k_y \sigma_1)-2t(\cos k_x+\cos k_y)-\mu$, with $t$, $A$ and $\mu$ being hopping amplitude, RSOC strength and chemical potential respectively, and Pauli matrices $\sigma_{1,2,3}$, $\tau_{1,2,3}$ act in spin and Nambu space separately. The superconducting term $\Delta_s(\bm k) = \Delta_0+2\Delta_1(\cos k_x+\cos k_y)$ in Eq. (\ref{eq:H}), denoting extended $s$-wave pairing, and $\Delta_{sd}(\bm k) = -2\Delta_2(\cos k_x+\delta \cos k_y +\eta)$, describing $s+d$ pairing that in addition exhibits a $\pi/2$-phase shift relative to $\Delta_s$. This time-reversal-symmetry(TRS)-broken pairing reduces to $s+d$ when $\Delta_s =0$, to $s+id$ when $\delta=-1$, $\eta=0$, and to $s+is$ when $\delta=1$. The energy spectrum of Hamiltonian Eq. (\ref{eq:H}) has a simple form,
\begin{equation}
E(\bm k) = \pm \sqrt{\epsilon_\pm^2(\bm k)+\Delta_s^2(\bm k)+\Delta_{sd}^2(\bm k)},\label{eq:spectrum}
\end{equation}
where $\epsilon_\pm(\bm k) = \pm 2A\sqrt{\sin^2 k_x+\sin^2 k_y}-2t(\cos k_x+\cos k_y)-\mu$, being the kinetic energy.

In the absence of $\Delta_{sd}$ term, the model is well known to support first-order topological superconducting phase that features TRS-protected helical Majorana modes on the edges, as well as nodal superconducting phase with point nodes \cite{Zhang2013b} (see Fig. \ref{fig1}(c)), provided  
\begin{equation}
|\mu - 4t\alpha_\Delta| < 2\sqrt{2}|A|\sqrt{1-\alpha_\Delta^2},\ \ |\alpha_\Delta|<1 \label{eq:c2}
\end{equation}
with $\alpha_\Delta=\Delta_0/(4\Delta_1)$. Equation (\ref{eq:c2}) can be fulfilled when the system exhibits $s_\pm$ pairing symmetries. Turning on $\Delta_{sd}$ is supposed to break TRS and gap out the helical modes. Instead of driving the system into trivial phases, we will demonstrate that this TRS-broken term may give birth to second-order topological superconducting phases, featuring MZMs bound at corners. To understand the origin of second-order phases, we may start from gapless edge states in the absence of $\Delta_{sd}$ and then consider effects of this mass term on the gapless modes.

As is known, second-order phases appear when gapless states on intersecting edges acquire mass gaps of opposite signs. To investigate the edge physics, we consider a cylinder geometry, where the periodic boundary condition is only assumed along the $y$ direction [see Fig. \ref{fig1}(a)]. Accordingly, the Hamiltonian in this geometry would take the form $H = \frac{1}{2}\sum_{k_y}\Psi^\dagger(k_y)\mathcal H_{\text{1D}}(k_y)\Psi(k_y)$, when written in the new basis $\Psi(k_y) = \oplus_j \psi_j(k_y)$, where $\psi_j(k_y) = \{c_{j,k_y\uparrow}, c_{j,k_y\downarrow}, c^\dagger_{j,-k_y\downarrow}, -c^\dagger_{j,-k_y\uparrow}\}^T$, $c_{j,k_y\uparrow(\downarrow)}= \frac{1}{\sqrt{N_x}}\sum_{k_x} c_{\bm k\uparrow(\downarrow)} e^{ik_xj}$, $j$ stands for lattice site and $N_x$ is the total number of sites. The components of $\mathcal H_{\text{1D}}(k_y)$ are given by the following $4\times 4$ block matrices,
\begin{eqnarray}
&&[\mathcal H_{\text{1D}}(k_y)]_{j, j} = \mathbf{M} = M^{\alpha\beta} \Gamma_{\alpha\beta},\label{eq:H1d}\\\nonumber
&&[\mathcal H_{\text{1D}}(k_y)]_{j, j+1} = ([\mathcal H_{\text{1D}}(k_y)]_{j+1, j})^\dagger =  \mathbf{T} = T^{\alpha\beta}\Gamma_{\alpha\beta}.\label{eq:H_cy}
\end{eqnarray}
In Eq. (\ref{eq:H1d}) $\Gamma_{\alpha\beta} = \tau_\alpha\otimes\sigma_\beta$ with $\alpha, \beta = 0, 1, 2, 3$, and the two tensors $M$ and $T$ have the following entries: $M^{30} = -\mu-2t \cos k_y$, $M^{31} = -2A \sin k_y$, $M^{10} = \Delta_0+2\Delta_1\cos k_y$, $M^{20} = -2\Delta_2(\eta+\delta\cos k_y)$, $T^{30} = -t$, $T^{32} = -iA$, $T^{10} = \Delta_1$ and $T^{20} = -\Delta_2$. The energy spectrum and corresponding wave functions in this geometry could be determined from the eigenvalue equation $\mathcal H_{\text{1D}}(k_y)\phi=E(k_y)\phi$, which leads to
\begin{equation}
 \mathbf{M}\phi_j+ \mathbf{T}^\dagger\phi_{j-1}+ \mathbf{T}\phi_{j+1} = E(k_y)\phi_j, \text{ for any } j, \label{eq:eigen1d}
\end{equation}
with $\phi_j$ being a four-component vector that represents the wave function at site $j$.

\begin{figure}[b]
\includegraphics[scale=0.32]{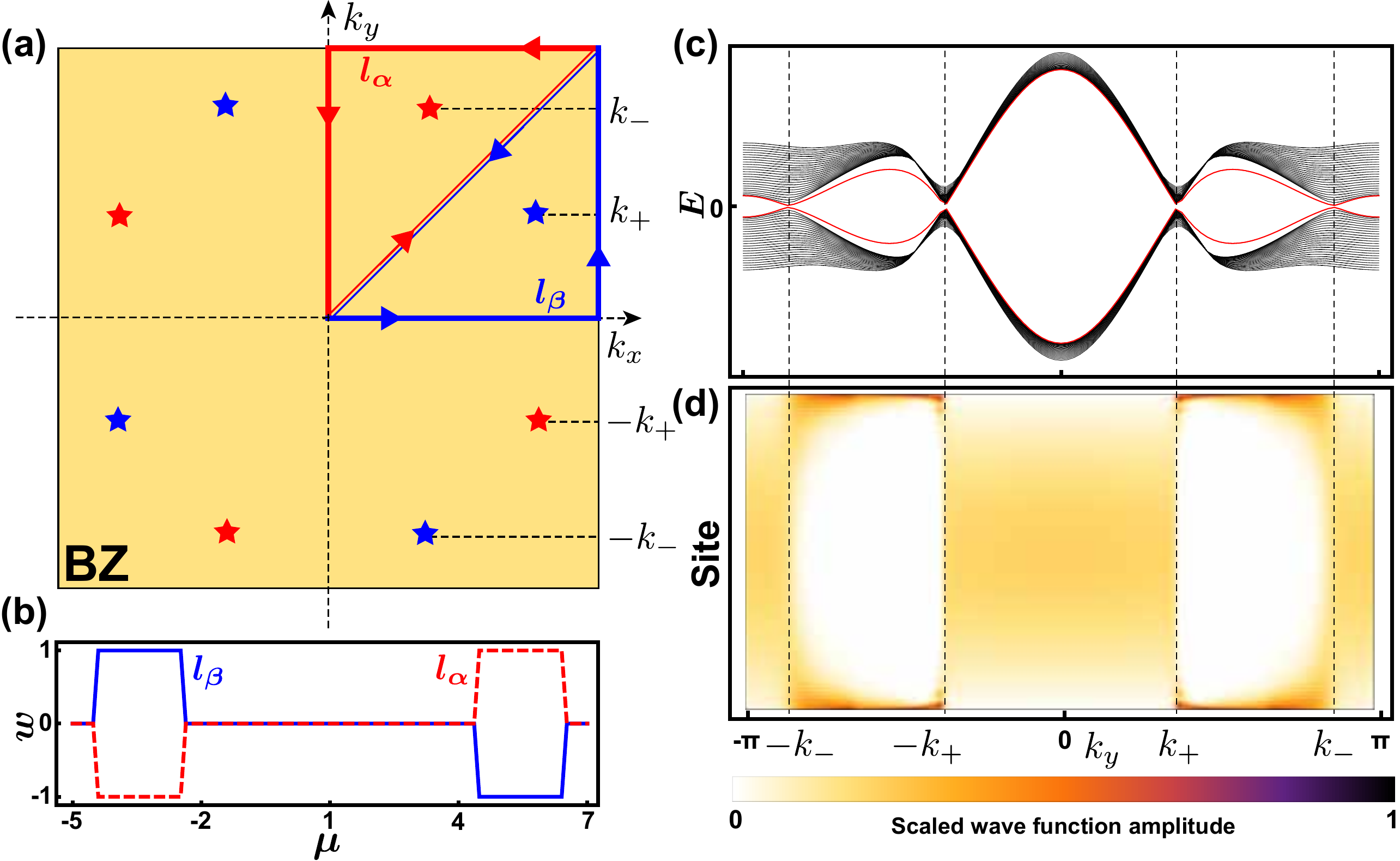}
\caption{Edge states in the nodal phase. (a) Bulk nodes in BZ are denoted by stars, and those in the same color have the same topological charge. (b) Winding number for topologically distinct nodes and its evolution with chemical potential. Clearly, only in the nodal phase would the winding number take nonzero values. (c) Energy spectrum in the cylinder geometry. Red lines denote the four energy levels closest to zero and each of the lines is doubly degenerate. (d) Variations of the (scaled) wave function amplitude $|\phi_j|$ with lattice site $j$ and wave vector $k_y$ for the energy levels denoted by red lines in (c). The parameters chosen are, $N_x=200$, $A=2$, $\Delta_0=\Delta_1=1$, $\Delta_2=0$. In (c) and (d), $\mu=5$.}\label{fig2}
\end{figure}

In the first-order phase when $\Delta_{sd}=0$, we have $E(k_y=\pi)=0$ if $\alpha_\Delta>0$ and $E(k_y=0)=0$ otherwise \cite{Zhang2013b}. In both cases MZMs are doubly degenerate on edge $AB$ as well as $CD$ defined in Fig. \ref{fig1}(a). Without loss of generality, hereafter we will assume $\alpha_\Delta>0$. In the nodal phase, zero modes in the spectrum $E(k_y)$ would appear at the projections of bulk nodes on the edge Brillouin zone (BZ), as is shown in Figs. \ref{fig2}(a) and \ref{fig2}(c). There are eight nodes in total, which relate to one another through fourfold rotation $\mathcal C_4$, mirror reflections $\mathcal M_x$ and $\mathcal M_y$ with mirror planes at $k_x=0$ and $k_y=0$. In the absence of $\Delta_{sd}$, Hamiltonian Eq. (\ref{eq:H}) is invariant under these operations, i.e.,
\begin{eqnarray}
&&\mathcal U_{\mathcal C_4}^{-1} \mathcal H(k_x, k_y)\mathcal U_{\mathcal C_4} = \mathcal H(-k_y, k_x),\nonumber\\
&&\mathcal U_{\mathcal M_x}^{-1} \mathcal H(k_x, k_y)\mathcal U_{\mathcal M_x} = \mathcal H(-k_x, k_y),\nonumber\\
&&\mathcal U_{\mathcal M_y}^{-1} \mathcal H(k_x, k_y)\mathcal U_{\mathcal M_y} = \mathcal H(k_x, -k_y),\label{eq:symmetry}
\end{eqnarray}
where $\mathcal U_{\mathcal C_4} = e^{i\pi\sigma_3/4}$, $\mathcal U_{\mathcal M_x } = \sigma_1$, and $\mathcal U_{\mathcal M_y} = \sigma_2$. Because of these crystalline symmetries, we may denote the eight bulk nodes by $\pm (k_+,\pm k_-)$ and $\pm(k_-, \pm k_+)$, with 
\begin{equation}
\cos k_\pm = -\alpha_\Delta \pm\sqrt{1-\alpha_\Delta^2-(\mu-4t\alpha_\Delta)^2/(8A^2)}.\label{eq:nodek}
\end{equation}
In contrast to the first-order phase, zero modes at $\pm k_\pm$ in the nodal phase are not localized. However, in the edge BZ where $k_y\in (-k_-,-k_+)\cup(k_+,k_-)$ ($0<k_+<k_-<\pi$ is assumed), we find that two localized states with opposite excitation energy $\pm E(k_y)$ would exist on each edge, as is evidenced in Figs. \ref{fig2}(c) and \ref{fig2}(d). It seems that these edge states are not topologically protected, since each gapless point ($\pm k_\pm$) is the projection of two bulk nodes carrying opposite topological charges [see Fig. \ref{fig2}(a)] which are supposed to cancel each other out. The charge for each bulk node is defined by the winding number $w$ along a contour $l$ surrounding this node \cite{Schnyder2011}, as is shown in Fig. \ref{fig2}(a) (see Supplemental Material for details). Possibly, these localized edge states are the remnants of those in the first-order phase. In our specific model defined in Eq. (\ref{eq:H}), they are robust provided the system is in the nodal phase. Hence we may describe the low-energy physics of each edge with a gapless Hamiltonian that is defined only at $k_y\in (-k_-,-k_+)\cup(k_+,k_-)$.

So we have established that MZMs emerge in both the first-order and the nodal phase when $\Delta_{sd}=0$. Edge states in these two phases could be well described by a one-dimensional massless Hamiltonian, with gapless points at $k_y^c = \pi$ in the first-order phase, and at $k_y^c = \pm k_\pm$ in the nodal phase. Note that Hamiltonian Eq.  (\ref{eq:H_cy}) preserves chiral symmetry $\Gamma_{20}$ in the absence of $\Delta_{sd}$, which guarantees that, for any state $\phi$ with finite energy $E(k_y)$ there would be a state $\Gamma_{20}\phi$ (shorthand for $\oplus_j\Gamma_{20}\phi_j$) with opposite energy $-E(k_y)$. Hence we can define the MZM basis for each edge as $\{\phi(k_y^c), \Gamma_{20}\phi(k_y^c)\}^T$. Instead of going into the details of MZMs, we will attempt to construct an effective edge Hamiltonian with a unified form.

First, multiplying Eq. (\ref{eq:eigen1d}) with $\phi_j^\dagger \Gamma_{10}$ on both sides, summing over $j$ and then adding to it the Hermitian conjugating counterpart, we are then left with
\begin{equation}
M^{10}=\sum\limits_j E(k_y)\phi_j^\dagger \Gamma_{10}\phi_j-T^{10}\phi_j^\dagger (\phi_{j-1}+\phi_{j+1}),\label{eq:m1}
\end{equation}
where the normalization condition $\phi^\dagger\phi=1$ is used. One could also multiply Eq. (\ref{eq:eigen1d}) with $\phi_j^\dagger \Gamma_{20}\Gamma_{10}$, and follow the same procedure as above, which would lead to
\begin{equation}
T^{10}\sum\limits_j \phi_j^\dagger\Gamma_{20}( \phi_{j-1}+\phi_{j+1})= 0,\label{eq:m2}
\end{equation}
due to orthogonality condition $\phi^\dagger\Gamma_{20}\phi=0$. At the gapless point $k_y^c$, Eq. (\ref{eq:m1}) reduces to
\begin{equation}
M^{10}=-T^{10}\sum\limits_j\phi_j^\dagger (\phi_{j-1}+\phi_{j+1}).\label{eq:m3}
\end{equation}
After projecting Hamiltonian Eq. (\ref{eq:H1d}) onto the MZM basis and utilizing the two equalities in Eqs. (\ref{eq:m2}) and (\ref{eq:m3}), one arrives at the effective low-energy Hamiltonian for edge $AB$ or $CD$, given by
\begin{equation}
\mathcal H_{\text{Edge}}(k_y) = v_2(k_y) s_2 + v_3(k_y) s_3 + m_{sd}(k_y) s_1,\label{eq:edgeH1}
\end{equation}
where 
\begin{eqnarray}
&&v_2(k_y) = \sum\limits_{j, \{\alpha\beta\}}[M^{\alpha\beta}(k_y)-M^{\alpha\beta}(k_y^c)]\phi_j^\dagger\Gamma_{20}\Gamma_{\alpha\beta}\phi_j,\nonumber\\
&&v_3(k_y) = \sum\limits_{j, \{\alpha\beta\}}[M^{\alpha\beta}(k_y)-M^{\alpha\beta}(k_y^c)]\phi_j^\dagger\Gamma_{\alpha\beta}\phi_j, \nonumber\\
&&m_{sd}(k_y) = -2\Delta_2(\delta\cos k_y+\eta-\cos k_y^c - 2\alpha_\Delta),\label{eq:gap}
\end{eqnarray}
with indices $\{\alpha\beta\}$ taking $\{30, 31, 10\}$ and Pauli matrices $s_{1,2,3}$ acting in the MZM basis. Wave functions of MZMs --- $\phi_j$ in Eq. (\ref{eq:gap}) --- could be obtained by solving Eq. (\ref{eq:eigen1d}) in principle, although we don't have to, given that it is the mass gap $m_{sd}$ that we care foremost, and that it clearly doesn't depend on the specific form of $\phi_j$. With the edge Hamiltonian Eq.   (\ref{eq:edgeH1}) being given, the condition when second-order phases emerge can be determined by comparing signs of mass gaps on intersecting edges, which we shall detail in the following.

\begin{figure}[h]
\includegraphics[scale=0.3]{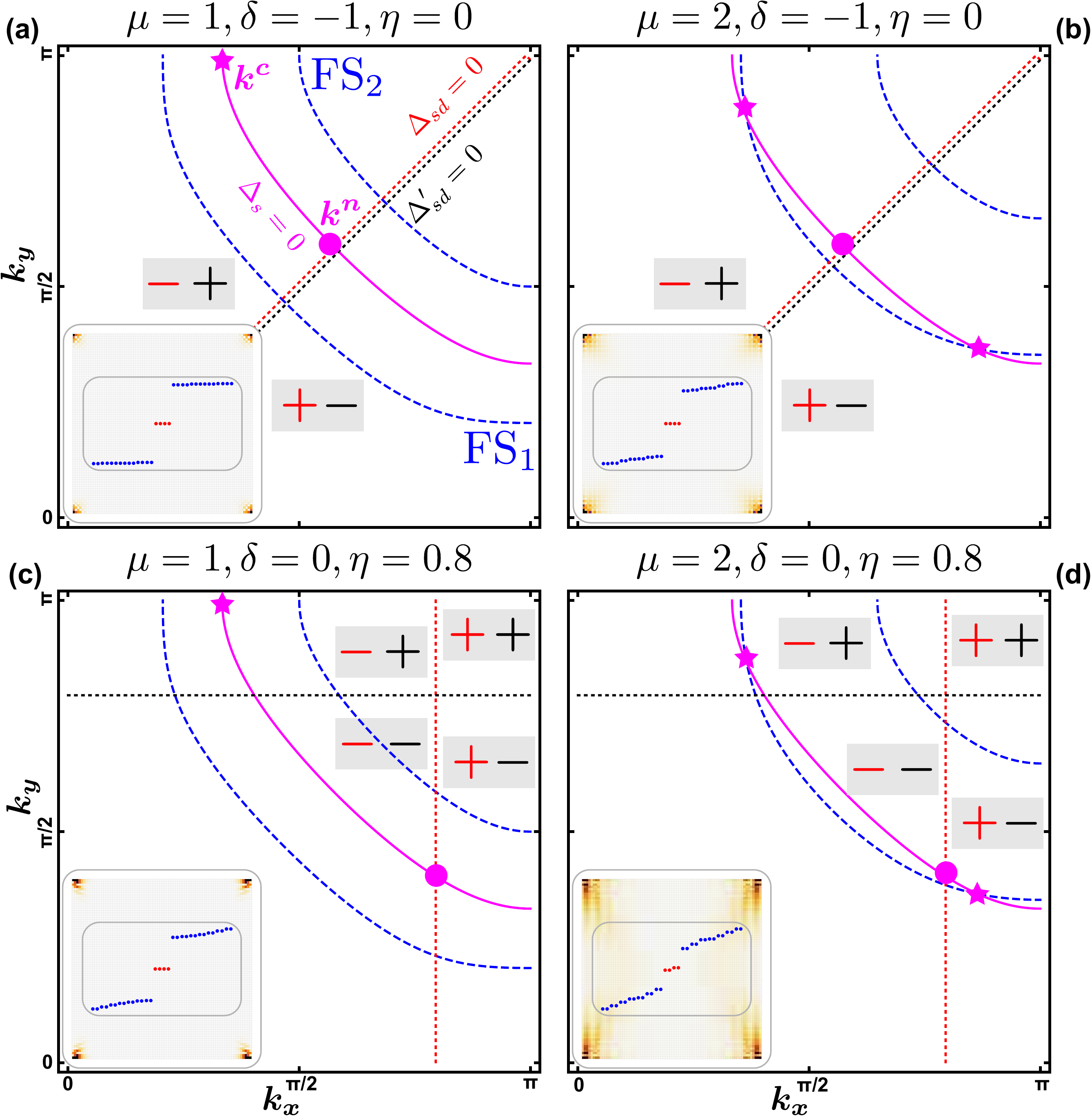}
\caption{Determination of the second-order phase from the bulk spectrum. (a)-(d) Fermi surfaces $\epsilon_\pm=0$, nodal lines of $\Delta_s$, $\Delta_{sd}$ and $\Delta_{sd}'$ in the first quadrant of BZ are plotted for different chemical potential ($\mu$) and $s+d$ pairing form ($\delta$, $\eta$). Signs of the pair $[\Delta_{sd}(\bm k), \Delta_{sd}'(\bm k)]$ are indicated in corresponding areas. The system resides in second-order phases when the signs of $\Delta_{sd}$ and $\Delta_{sd}'$ at $\bm{k^c}$ (marked by stars in magenta) are opposite, or equivalently, when the nodal point $\bm{k^n}$ (marked by magenta circle) of the pairing term lies between the two Fermi surfaces. Distributions of MZMs for an $80\times 80$ lattice are shown in the insets, as well as several low-lying energy levels. Clearly, MZMs (red points in the insets) are fourfold degenerate and separated from other energy levels with a finite gap. In all the figures, $A=\Delta_2=0.5$, $\Delta_0=\Delta_1=1$.}\label{fig3}
\end{figure}

Let us consider rotating the basis in Eq. (\ref{eq:H}) to $\Psi'(\bm k') = \mathcal U_{\mathcal C_4}\Psi(\mathcal C_4\bm k')$, where $\bm k'$ stands for coordinates in the $O-k_x'k_y'$ system defined in Fig. \ref{fig1}(a) and relates to $\bm k$ through $\mathcal C_4$ rotation $\mathcal C_4 \bm k' = \bm k$, namely, $(-k_y', k_x')=(k_x, k_y)$. Rewriting Hamiltonian Eq.  (\ref{eq:H}) in this new basis, we would have
\begin{eqnarray}
&&H = \frac{1}{2}\sum\limits_{\bm k'}\Psi'^{\dagger} (\bm k') \mathcal H'(\bm k') \Psi'(\bm k'),\label{eq:H1}\\
&&\mathcal H'(\bm k') = \mathcal U_{\mathcal C_4} \mathcal H(\mathcal C_4\bm k') \mathcal U_{\mathcal C_4}^{-1} = h(\bm k')+\Delta_s(\bm k')+\Delta_{sd}'(\bm k'),\nonumber
\end{eqnarray}
where $\Delta_{sd}'(\bm k')=-2\Delta_2(\cos k_y'+\delta \cos k_x' +\eta)$ and the last equality in Eq. (\ref{eq:H1}) is due to $\mathcal C_4$ symmetry of $h$ and $\Delta_s$ detailed in Eq. (\ref{eq:symmetry}). Comparing the two Hamiltonians in Eqs.(\ref{eq:H}) and (\ref{eq:H1}), one may immediately conclude that the edge Hamiltonian along edge $AD$ or $BC$ could be obtained from Eq. (\ref{eq:edgeH1}) simply by replacing $k_y$ with $k_y'$, followed by modification of the mass term, which yields
\begin{equation}
\mathcal H_{\text{Edge}}'(k_y') = v_2(k_y') s_2 + v_3(k_y') s_3 + m_{sd}'(k_y') s_1,\label{eq:edgeH11}
\end{equation}
with
\begin{equation}
m_{sd}'(k_y') = -2\Delta_2(\cos k_y'+\eta - \delta \cos k_y^{c} - 2\delta\alpha_\Delta),\label{eq:gap1}
\end{equation}
and the definitions of $v_2$ and $v_3$ are given in Eq. (\ref{eq:gap}). It is obvious that gapless points in the two edge Hamiltonian, $\mathcal H_{\text{Edge}}(k_y)$ and $\mathcal H_{\text{Edge}}'(k_y')$, both reside at $k_y^c$. The second-order phase therefore emerges when
\begin{equation}
m_{sd}(k_y^c) m_{sd}'(k_y^c) < 0. \label{eq:c1}
\end{equation}
Additionally, we require Eq. (\ref{eq:c2}) to be fulfilled, which guarantees that the system falls into the first-order or nodal phase when $\Delta_{sd}$ is switched off.

Further investigations on Eqs. (\ref{eq:gap}) and (\ref{eq:gap1}) reveal that, the mass terms $m_{sd}(k_y^c)$ and $m_{sd}'(k_y^c)$ are nothing but values of $\Delta_{sd}(\bm k)$ and $\Delta_{sd}'(\bm k')$ at point $\bm{k^c}=(k_x^c, k_y^c)$ that satisfies $\Delta_s(\bm{k^c}) = 0$, with $k_y^c$ being the gapless point in the edge BZ. Thus we may relate the criterion obtained from the edge Hamiltonian with the bulk spectrum in Eq. (\ref{eq:spectrum}). As illustrated in Fig. \ref{fig3}, Eq. (\ref{eq:c1}) actually requires $\Delta_{sd}$ and $\Delta_{sd}'$ to take opposite signs at $\bm{k^c}$ marked by stars, that is,
\begin{equation}
\Delta_{sd}(\bm{k^c})\Delta_{sd}'(\bm{k^c})<0.
\end{equation}
Substituting the expression of $k_y^c$ into Eq. (\ref{eq:c1}), we arrive at the conditions for second-order phases,
\begin{eqnarray}
&&|\eta-f_1|<|f_3|, \label{eq:c32}\\
&&|\mu-4t\alpha_\Delta|<|\frac{2\sqrt{2}A}{1-\delta}|\sqrt{f_2^2-(\eta-f_1)^2},\label{eq:c31}
\end{eqnarray}
with $f_1=(1+\delta)\alpha_\Delta$, $f_2 = (1-\delta)\sqrt{1-\alpha_\Delta^2}$ and $f_3 = (1-\delta)(1-\alpha_\Delta)$. Equation (\ref{eq:c32}) determines which kind of pairing form could possibly induce the second-order phase, while Eq. (\ref{eq:c31}) establishes the relation of Fermi surfaces with the pairing potential in this nontrivial phase. Indeed, we observe that the nodal point $\bm{k^n}$ ($\Delta_s(\bm{k^n}) = \Delta_{sd}(\bm{k^n})=0$) of the superconducting order parameter, marked by a magenta circle in Fig. \ref{fig3}, always lies between the two Fermi surfaces in the second-order phase. This is verified by the fact that Eq. (\ref{eq:c31}) could also be obtained by requiring 
\begin{equation}
\epsilon_+(\bm{k^n})\epsilon_-(\bm{k^n})<0,\label{eq:c4}
\end{equation}
where $\epsilon_\pm$ are the same as those in Eq. (\ref{eq:spectrum}) and take zero separately on the two Fermi surfaces. In addition, we also note that Eq. (\ref{eq:c32}) actually guarantees the existence of nodal point $\bm{k^n}$. Therefore, one may determine when the system resides in the second-order phase, either from Eqs. (\ref{eq:c32}) and (\ref{eq:c31}), or from Eq.  (\ref{eq:c4}), as illustrated in Fig. \ref{fig3}. Following these criteria, one may immediately conclude that $s+id$ pairing favors the second-order phase while neither $s+d$ nor $s+is$ pairing do. 

The mixed-pairing state we consider above has been extensively studied in iron pnictides, particularly 122 compounds \cite{Tafti2013,Kretzschmar2013,Bohm2014,Tafti2015,Guguchia2015,Guguchia2016,Li2017nc} like Ba$_{1-x}$K$_x$Fe$_2$As$_2$. In these materials, the pairing symmetry is expected to change from a nodeless $s_\pm$ form around optimal doping ($x \sim 0.4$) \cite{Hosono2015} to a form with nodal gaps in the heavily hole-doped region, for instance, KFe$_2$As$_2$ ($x = 1$). In a narrow doping region between the two cases, two different pairing states may coexist with an additional $\pi/2$-phase shift at lower temperature when TRS is spontaneously breaking  \cite{Khodas2012,Guguchia2016,Platt2017}. The main debate, however, centers around the heavily hole-doped region, where multiple experiments suggest contradicting pairing, either nodal $s$- \cite{Okazaki2012,Cho2016} or $d$ wave \cite{Tafti2013,Tafti2015,Hafiez2013,Grinenko2014}. Accordingly, the intermediate state would exhibit either $s+is$ or $s+id$ symmetry, as was reported in muon spin rotation experiment at doping level around $x = 0.73$ \cite{Grinenko2017}. No consensus has been achieved as to which specific form it would take, although several proposals have been put forward to discriminate the two mixed pairings \cite{Maiti2015,Lin2016,Garaud2016}. In this regard, our study suggests an alternative approach to tackle this issue, given that $s+id$ could drive a RS into the second-order phase with MZMs sitting at corners whereas $s+is$ pairing couldn't. Once $s+id$ pairing has been confirmed, it would be straightforward to fabricate the heterostructure as depicted in Fig. \ref{fig1}(a) and to investigate MZMs being expected therein.

Meanwhile, we may also consider a hybrid Josephson junction, as schematically shown in Fig. \ref{fig1}(b). The FeSC on top with $s_\pm$ pairing and the cuprate SC at the bottom with $d_{x^2-y^2}$ pairing may introduce a mixed state of the form $s+e^{i\theta}d$ in the RS layer sandwiched between them. In the Supplemental Material we demonstrate that this kind of pairing falls into the general form studied above, and interestingly it always favors second-order phases except when the phase difference $\theta$ of the two SCs takes 0 or $\pi$, which corresponds to $s\pm d$ pairing. To guarantee that the phase difference never takes 0 or $\pi$, one may insert this hybrid system into a single-junction rf SQUID \cite{Komissinski2002} or a two-junction dc SQUID \cite{Katase2010}, where $\theta$ may be tuned through magnetic flux threaded into the interferometer. Actually, it has been suggested that such a hybrid system could naturally realize a junction with $\theta=\pi/2$ and thus $s+id$ pairing order would develop at the interface \cite{Yang2018}. Hybrid Josephson junctions containing conventional $s$-wave SCs and FeSCs \cite{Zhang2009,Kalenyuk2018} or cuprate SCs \cite{Kleiner1996,Komissinskiy2007} have been successfully fabricated and well studied. We can therefore expect the hybrid junction with an FeSC, RS, and cuprate SC to be a promising platform for second-order TSCs in the near future. 

The author acknowledges J. Liu for many illuminating discussions. This work was supported by National Natural Science Foundation of China (NSFC) under Grant No. 11704305 and No. 11847236.

\bibliography{mcssd.bib}

\clearpage

\appendix
\setcounter{equation}{0}
\setcounter{page}{1}

\section{Supplemental Material for ``Second-order topological superconductors with mixed pairing"}

\subsection{A. Winding number in the nodal phase}

In the nodal superconducting phase of our two-dimensional system, there are eight point nodes sitting in the Brillouin zone. For each node one can define the topological charge as the winding number of a closed loop that encircles only this particular gapless point. Additionally, we require the loop not to cross any other node, so that the winding number can be well defined over the loop. In the following we shall illustrate how to calculate the winding number and corresponding topological charge of the point node for our given system (one may refer to Ref. \cite{Schnyder2011} and the supplemental material therein for a detailed and general analysis). 

Note that in the absence of $\Delta_{sd}$ term, our model Hamiltonian reduces to 
\begin{equation}
\mathcal H(\bm k) = h(\bm k)\tau_3+\Delta_s(\bm k)\tau_1.
\end{equation}
Owing to the chiral symmetry $\mathcal S = \tau_2$, this Hamiltonian can be brought into block off-diagonal form, which reads
\begin{equation}
\tilde{\mathcal H}(\bm k) = \mathcal U^{-1} \mathcal H(\bm k)\mathcal U = \begin{pmatrix} 0 & D(\bm k) \\ D^\dagger(\bm k) & 0\end{pmatrix},\label{eq:H_od}
\end{equation}
with $\mathcal U = e^{i\frac{\pi}{3}\cdot\frac{1}{\sqrt 3}(\tau_1+\tau_2+\tau_3)}$, and $D(\bm k) = h(\bm k)-i\Delta_s(\bm k) \mathbb 1_2$, being a $2\times 2$ complex matrix. Since the loop we choose doesn't cross any node, energy spectrum is thus gapped everywhere on it. In this case, it would be very convenient to work with a spectrally flatten Hamiltonian while calculating the winding number over a given loop. The spectrally flatten Hamiltonian takes the similar form as in Eq. (\ref{eq:H_od}), being
\begin{equation}
Q(\bm k) = \begin{pmatrix} 0 & q(\bm k) \\ q^\dagger(\bm k) & 0\end{pmatrix}, \ \ q(\bm k)=U(\bm k)V^\dagger(\bm k).\label{eq:Q}
\end{equation}
$U(\bm k)$ and $V^\dagger(\bm k)$ in Eq.  (\ref{eq:Q}) are $2\times 2$ unitary matrices and relate to $D(\bm k)$ through singular-value decomposition,
\begin{equation}
D(\bm k) = U(\bm k)\tilde D(\bm k)V^\dagger(\bm k),\label{eq:svd}
\end{equation}
with $\tilde D(\bm k)$ being a diagonal matrix where all the entries are real and positive (note that $\det[D(\bm k)]\neq 0$ for all $\bm k$ on the loop). Clearly, $q(\bm k)$ is also a $2\times 2$ unitary matrix and is well defined on the two-dimensional BZ except at the eight point nodes. 

In general, $U$ and $V$ may be expressed using eigenstates of $D^\dagger D$ or $DD^\dagger$. Denote the eigenstates of $D^\dagger D$ by $\varphi_n$, with corresponding eigenvalues being $\xi_n$, and we have
\begin{equation}
D^\dagger D\varphi_n=\xi_n\varphi_n, \ \ n = 1, 2.\label{eq:eigen_D}
\end{equation}
After subsitituting Eq. (\ref{eq:svd}) into Eq. (\ref{eq:eigen_D}), we arrive at
\begin{equation}
\tilde D^2V^\dagger\varphi_n=\xi_nV^\dagger\varphi_n.
\end{equation}
Apparently, $\xi_n$ are the eigenvalues of diagonal matrix $\tilde D^2$ and $V^\dagger\varphi_n$ the eigenstates. As a result, $\tilde D^2$ can be written as 
\begin{equation}
\tilde D^2 = \begin{pmatrix} \xi_1 & 0 \\ 0 & \xi_2\end{pmatrix},
\end{equation}
and we may choose 
\begin{equation}
V^\dagger\varphi_1 =  \begin{pmatrix} 1\\ 0\end{pmatrix}, \ \ V^\dagger\varphi_2 = \begin{pmatrix} 0\\ 1\end{pmatrix},
\end{equation}
that is,
\begin{equation}
V^\dagger \begin{pmatrix} \varphi_1 & \varphi_2 \end{pmatrix} = \mathbb 1_2.
\end{equation}
Due to the orthogonality and normalization of $\varphi_n$, we thus have
\begin{equation}
V = \begin{pmatrix} \varphi_1 & \varphi_2 \end{pmatrix}, \ \ V^\dagger = \begin{pmatrix} \varphi_1^\dagger \\ \varphi_2^\dagger \end{pmatrix}.
\end{equation}
The unitary matrix $U$ can be obtained using the relation in Eq. (\ref{eq:svd}), and takes the form
\begin{equation}
U = DV\tilde D^{-1} = \begin{pmatrix} \xi_1^{-1}D\varphi_1 & \xi_2^{-1}D\varphi_2 \end{pmatrix}.
\end{equation}
The $q$-matrix can then be readily obtained, which reads
\begin{equation}
q(\bm k) = UV^\dagger = \xi_1^{-1}D\varphi_1\varphi_1^\dagger+\xi_2^{-1}D\varphi_2\varphi_2^\dagger. \label{eq:qk}
\end{equation}
Alternatively, one may also express $q$-matrix with eigenstates $\tilde\varphi_n$ of $DD^\dagger$, where 
\begin{equation}
DD^\dagger\tilde\varphi_n = \xi_n\tilde\varphi_n.
\end{equation}
Following the same procedure as above, we would have
\begin{equation}
q(\bm k) = UV^\dagger = \xi_1^{-1}\tilde\varphi_1\tilde\varphi_1^\dagger D+\xi_2^{-1}\tilde\varphi_2\tilde\varphi_2^\dagger D.
\end{equation}
Note that there is a freedom in multiplying $\varphi_n (\tilde\varphi_n)$ by a phase factor $e^{i\theta_n}$ when $\xi_1\neq\xi_2$, which obviously doesn't alter the form of $q(\bm k)$. If the spectrum is degenerate, \emph{i.e.}, $\xi_1=\xi_2 = \xi$, Eq. (\ref{eq:qk}) would simply reduce to
\begin{equation}
q(\bm k) = \xi^{-1}D(\bm k),
\end{equation}
from which it follows immediately that $q(\bm k)$ is invariant under rotation of the orthonormal basis $\{\varphi_1, \varphi_2\}$ in the linear space spanned by them.

For the spectrally flatten Hamiltonian in Eq. (\ref{eq:Q}), the winding number over a given loop $l$ is simply given by
\begin{equation}
w = \frac{1}{2\pi i}\oint_l dl\ \text{Tr}[q^{-1}(\bm k) \nabla_l q(\bm k)].\label{eq:winding}
\end{equation}
$w$ exactly defines the topological charge of a point node when the path $l$ is chosen to be circling around this particular node only. It should be noted that the integral in Eq. (\ref{eq:winding}) is taken to be counterclockwise along path $l$ in the main text.

\subsection{B. Phase difference between $s_\pm$ and $d_{x^2-y^2}$ pairing in hybrid Josephson junction}

Consider an arbitrary phase difference $\theta$ between the iron-based SC  ($s_\pm$ pairing) and cuprate SC ($d_{x^2-y^2}$ pairing) in hybrid Josephson junction depicted in the main text. Rashba layer sandwiched between the two SCs would develop a superconducting term with mixed pairing, given by
\begin{equation}
(\tilde\Delta_s + e^{i\theta}\tilde\Delta_d)(c^\dagger_{\bm k\uparrow}c^\dagger_{-\bm k\downarrow}-c^\dagger_{\bm k\downarrow}c^\dagger_{-\bm k\uparrow})+\text{H.c.},\label{eq:pairing}
\end{equation}
where
\begin{eqnarray}
\tilde\Delta_s &&= \tilde\Delta_0 + 2\tilde\Delta_1(\cos k_x+\cos k_y), \\
\tilde\Delta_d &&= 2\tilde\Delta_2(\cos k_x-\cos k_y).
\end{eqnarray}
We may perform a gauge transformation on the Nambu spinor basis, which send $c_{\bm k}$ to 
\begin{equation}
\tilde c_{\bm k} = c_{\bm k}e^{-i\frac{1}{2}(\theta-\frac{\pi}{2})},
\end{equation}
and the pairing term in Eq. (\ref{eq:pairing}) could be rewritten as
\begin{equation}
[\tilde\Delta_s\sin\theta + i(\tilde\Delta_s\cos\theta+\tilde\Delta_d)](\tilde c^\dagger_{\bm k\uparrow}\tilde c^\dagger_{-\bm k\downarrow}-\tilde c^\dagger_{\bm k\downarrow}\tilde c^\dagger_{-\bm k\uparrow})+\text{H.c.} \label{eq:pairing1}
\end{equation}
Clearly, Eq. (\ref{eq:pairing1}) exactly describes $s+i(s+d)$ pairing as in the main text, where the first term $\tilde\Delta_s\sin\theta$ is known to be responsible for first-order or nodal superconducting phase, and the second one with a $\pi/2$-phase shift would open a finite gap on each edge. Define
\begin{eqnarray}
&&\Delta_0 = \tilde\Delta_0\sin\theta, \ \Delta_1 = \tilde\Delta_1\sin\theta,\ \Delta_2 = (\tilde\Delta_1\cos\theta+\tilde\Delta_2)\nonumber\\
&&\delta = \frac{\cos\theta-\tilde\beta_\Delta}{\cos\theta+\tilde\beta_\Delta}, \ \ \eta = \frac{2\tilde\alpha_\Delta\cos\theta}{\cos\theta+\tilde\beta_\Delta},\label{eq:definition}
\end{eqnarray}
with $\tilde\alpha_\Delta = \tilde\Delta_0/(4\tilde\Delta_1)$, $\tilde\beta_\Delta = \tilde\Delta_2/\tilde\Delta_1$, and we could then recover the same pairing form as in the main text, \emph{i.e.},
\begin{equation}
(\Delta_s - i\Delta_{sd})(c^\dagger_{\bm k\uparrow}c^\dagger_{-\bm k\downarrow}-c^\dagger_{\bm k\downarrow}c^\dagger_{-\bm k\uparrow})+\text{H.c.},
\end{equation}
with
\begin{eqnarray}
\Delta_s &&= \Delta_0 + 2\Delta_1(\cos k_x+\cos k_y), \label{eq:deltas}\\
\Delta_{sd} &&= -2\Delta_2(\cos k_x+\delta\cos k_y+\eta).
\end{eqnarray}
Hence, those criteria obtained before could be applied straightforwardly in this circumstance. The resulting condition for the second-order topological phase is then given by
\begin{equation}
|\mu - 4t\Delta| < 2\sqrt{2}|A|\sqrt{1-\Delta^2}, \ \ \theta\neq 0, \pi.\label{eq:criterion}
\end{equation}
The requirement of $\theta\neq 0, \pi$ in Eq. (\ref{eq:criterion}) guarantees that the pairing amplitude $\Delta_s$ in Eq. (\ref{eq:deltas}) is nonzero. Also, we should note that the inequality, $|\tilde\alpha_\Delta| < 1$, is always valid for $s_\pm$ pairing. This result implies that the criterion for second-order phase is independent of the phase difference $\theta$ if only the latter doesn't take $0$ or $\pi$. Therefore, we don't require the phase difference of the hybrid Josephson junction to be fine-tuned to certain values. Instead, it can take any value other than $0$ and $\pi$, where the latter is known to be $s\pm d$ pairing. 

A subtle issue of Eq. (\ref{eq:definition}) is that, $\delta$ and $\eta$ therein would be ill defined when $\cos\theta+\tilde\beta_\Delta=0$. In this case, the mass gaps of adjacent edges would be given by
\begin{eqnarray}
m_{sd} &&= -4\tilde\Delta_1\cos\theta(\cos k_y^c+\tilde\alpha_\Delta),\\
m_{sd}' &&= -4\tilde\Delta_1\cos\theta(-\cos k_y^c-\tilde\alpha_\Delta),
\end{eqnarray}
and hence we always have $m_{sd}m_{sd}'<0$, and the criterion in Eq. (\ref{eq:criterion}) would still be valid.\\\\

\end{document}